\documentclass[twocolumn,showpacs,preprintnumbers,APS]{revtex4}
\usepackage{graphicx}
\usepackage{amsmath}
\usepackage{amssymb}
\usepackage[latin1]{inputenc}
\usepackage{dcolumn}
\usepackage{bm}

\begin{document}

\title{Transport in superlattices on single layer graphene.}

\author{P. Burset$^1$, A. Levy Yeyati$^1$, L. Brey$^2$ and H. A. Fertig$^3$ }
\affiliation{$^1$Departamento de F\'{\i}sica Te\'orica de la Materia
Condensada C-V, UAM, E-28049 Madrid, Spain \\ 
$^2$Instituto de Ciencia de Materiales de Madrid, (CSIC), Cantoblanco, E-28049 Madrid, Spain \\ 
$^3$Department of Physics, Indiana University, Bloomington IN 47405}

\date{\today}


\begin{abstract}
We study transport in undoped graphene in the presence of a superlattice potential
both within a simple continuum model and using numerical tight-binding
calculations.  The continuum model demonstrates that the conductivity
of the system is primarily impacted by the velocity anisotropy that
the Dirac points of graphene develop due to the potential.  For
one-dimensional superlattice potentials, new Dirac points may be
generated, and the resulting conductivities can be approximately
described by the anisotropic conductivities associated with each
Dirac point.  Tight-binding calculations demonstrate that this simple
model is quantitatively correct for a single Dirac point, and that
it works qualitatively when there are multiple Dirac points.  Remarkably,
for a two dimensional potential which may be very strong but introduces
no anisotropy in the Dirac point, the conductivity of the system remains
essentially the same as when no external potential is present.
\end{abstract}

\maketitle

\section{\label{sec:intro} Introduction}
Graphene is one of the most
interesting electronic systems to become available in the last few
years \cite{Geim_2007,Castro_Neto_RMP}. Graphene is a two-dimensional arrangement of carbon atoms in a
triangular lattice with two atoms per unit cell.
In graphene, the electronic
low energy properties  are governed by a massless Dirac Hamiltonian
and the carriers moving in graphene have very interesting
properties: the electronic spectrum is linear in the wavevector, and
their states are chiral with respect the pseudospin defined by the
two atoms of the crystal  unit cell. These properties are
responsible for exotic effects, such as a half-integer
quantum Hall effect \cite{Novoselov_2005,Zhang_2005} and the Klein
paradox -- perfect transmission through potential
barriers \cite{Kastnelson_2006}.

The application of electric fields via nano gate geometries makes it
possible to subject the system to potentials varying on a short
length scale. Using these techniques, recently it has been possible
to study experimentally transport through $p$-$n$ junctions and
$p$-$n$-$p$ junctions in graphene
\cite{Huard_2007,Young_2009,Stander_2009,Velasco_2009,Russo_2009}.
Theoretically, there has also been much effort devoted to the study
of the spectra  and  the electronic transport through differently
doped regions
\cite{Pereira_2006,Cheianov_2007,Zhang_2008,Beenakker_2008,Arovas_2010}
whose behavior differs from that of conventional two-dimensional
electron gases.

A superlattice potential on top of graphene opens the possibility of
tailoring its band structure and modifying its transport
properties\cite{Vazquez_2008,Pletikosic_2009,Tivari_2009,Guinea_2010,Park_2008c}.
In particular in the case of a one dimensional superlattice
potential, the properties of the carriers  are extremely sensitive
to the amplitude $V_0$ and period $d$ of the superlattice. For a one
dimensional superlattice, the velocity of the carriers is highly
anisotropic \cite{Park_2008a,Park_2008b,Barbier_2008} and the number
of Dirac points at the Fermi energy can be altered by varying  the
product $V_0 d$ \cite{Brey_2009,Park_2009,Barbier_2010}. Moreover, when
the potential magnitude of the superlattice varies slowly in space,
the electronic spectra develops a Landau level spectrum \cite{Sun_2010}.
The effect of superlattice potentials due to external magnetic fields has also attracted a great deal of attention \cite{Martino_2009,Snyman_2009,Tan_2010,Martino_2011}.

Several groups have numerically studied electronic transport perpendicular 
to the superlattice barriers \cite{Brey_2009,Abedpour_2009,Barbier_2009,Barbier_2010,Wang_2010,Yampolskii_2008,Bai_2007}.
Starting from the theoretical universal value $\sigma _0 =\frac 4
{\pi} \frac { e^2} h  $\cite{Beenakker_2008}, the conductivity
increases with the product $V_0d$ and develops peaks at the critical
values of $V_0d$ for which new Dirac points emerge\cite{Brey_2009}.

In this work we consider electronic transport in graphene in the
presence of superlattice potentials that are piecewise constant.
In the case of one-dimensional superlattices we study both transport
parallel [Fig. \ref{fig:esquema}(a)] and perpendicular [Fig. \ref{fig:esquema}(b)] to the barriers. We also analyze transport in
two-dimensional superlattices [Fig. \ref{fig:esquema}(c)].
Analytical expressions for the conductivity are obtained by describing
the carriers with the Dirac Hamiltonian and using the Kubo formula.
These are compared
with numerical results obtained using a tight-binding
Hamiltonian for graphene in the presence
of a superlattice potential and the Landauer-B\"{u}ttiker
formalism for obtaining the electrical conductivity in the presence of leads.

In the case of a one dimensional superlattice, we find that, as a
function of the product  $V_0d$, the conductivity parallel to the
superlattice barriers, $\sigma _\parallel$, decreases quadratically
from its value in the absence of the potential, $\sigma _0$, whereas
in the perpendicular direction the conductivity $\sigma _ \perp$
increases quadratically. The appearance of new Dirac points produces
peaks in $\sigma _ \perp$ and minima  in $\sigma _\parallel $. For
two-dimensional superlattices the conductivity depends on the
relative values of the product $V_0 d$ in different directions.
Interestingly, for isotropic superlattice potentials, the
conductivity is unaffected by the perturbation and remains at the
universal value $\sigma _0$=$\frac 4 {\pi} \frac { e^2} h $. Further insight into the character of transport is obtained from the channel decomposition of the transmission matrix.

This paper is organized as follows.  In Section II we present the analytical results for the conductivity obtained assuming
independent anisotropic Dirac points. In Section III we present numerical results obtained with a microscopic tight-binding Hamiltonian and compare with the analytical expressions. Section IV is dedicated to the conclusions.

\section{One dimensional superlattice potential}
\vspace{-0.3cm}
\subsubsection {Preliminaries.}  \vspace{-0.3cm}
The electronic structure of an infinitely large flat graphene flake is described by the Dirac Hamiltonian,
\begin{equation}
H _0 = \hbar v_F \textbf{k} \cdot \mathbf{\sigma }\, \, \,
\label{Dirac}
\end{equation}
where $\hbar \textbf{k}$ is the momentum operator, $\mathbf{\sigma}$
are the Pauli matrices and $v_F \simeq 10 ^6 m/s$ is the Fermi
velocity. The two entries of the Dirac Hamiltonian correspond to the
two carbon atoms in the  unit cell in graphene.

The eigenvalues and eigenvectors of this Hamiltonian are
$\varepsilon _{ k,s }$=$s v_F \hbar k$ and $|s,\mathbf{k}> = \frac
{e ^{i\mathbf{k} \mathbf{r}} }{\sqrt{2}} \left ( \begin{array}{c}
                                                          1 \\  s e ^{i \theta(\mathbf{k})}
                                                        \end{array}
                                                        \right )$,
where $s=-1$ and $s=1$ describe the occupied and empty bands
respectively. In the previous expressions $\theta (\mathbf{k})$ is
the angle of the vector $\mathbf{k}$ with respect to the $\hat{k}_x$
direction.

\vspace{-0.3cm}
\subsubsection {Superlattice band structure.}  \vspace{-0.3cm}
\begin{figure}[ht]
\includegraphics[width=8.5cm]{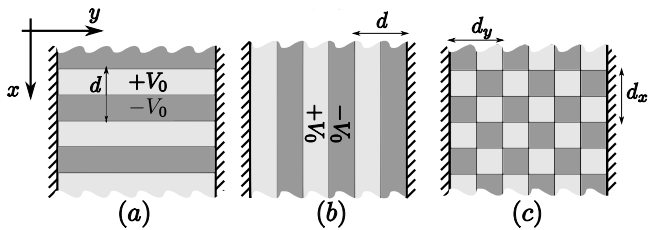}
\caption{(Color online) Schematic representation of the superlattices used showing the axis selection. The system is infinite along the $x$-direction and has a finite length along the $y$-direction. The dashed patterns on each side of the $y$-direction indicate the leads for the Landauer conductance calculations. The superlattice barriers can be parallel (a) or perpendicular (b) to the direction of transport. We also consider a chessboard-like two dimensional superlattice potential in which $d_y\simeq d_x$ (c).}
\label{fig:esquema}
\end{figure}

We consider a one-dimensional Kronig-Penney superlattice along the
$\hat{x}$-direction (see Fig. \ref{fig:esquema}(a)). The period of the potential  is $d$, $V(x)=V(x+d)$ and $V(x)$=$V_0 \, \textrm{sgn} (x)$ for $|x|<d/2$.
For this potential it is possible to find an analytical expression
for the band structure \cite{Barbier_2010,Arovas_2010}, that in the
limit of small wave vector and energies takes the form
\begin{equation}
\varepsilon (\mathbf{k})= \pm \hbar v_F \left ( k _x ^2 + k _y ^2 \,
\frac {\sin ^2 (\tilde{V})} { \tilde{V} ^2} \right ) ^{1/2} \, \, \,
, \label{bands_SL}
\end{equation}
where $\tilde{V}$=$\frac {V_0 d}{2 \hbar v_F}$. The group velocity
of the state is anisotropically renormalized, and has a strong
dependence on the direction of the wave vector $\mathbf{k}$\cite{Park_2008a}. At the Dirac point and for directions along the superlattice axis the velocity of the carriers is unaffected by the potentials, $v_x ^0$=$v_F$. However the group
velocity along the direction perpendicular to the superlattice direction is strongly renormalized and takes the form
\begin{equation} v_y ^0  \simeq v_F \frac {|\sin (\tilde{V})|} {\tilde{V}}
\, \, \, . \label{vy0}\end{equation}

Whenever the superlattice parameters satisfy the condition,
\begin{equation}
\frac {V_0 d} {\hbar v_F}= 2  \pi j \, \, \, \, \, j=1,2,3,... \, \,
\,\label{condic}
\end{equation}
the group velocity in the $\hat{y}$ direction vanishes and a new
pair of Dirac points emerges from the original Dirac point, moving in
opposite direction along the $\hat{k} _y$-direction\cite{Brey_2009,Park_2009}.
Near the new Dirac points and at low energy  the dispersion is also
linear and anisotropic. For the $j$-th pair of new Dirac points the
velocity in the $\hat{x}$ and $\hat{y}$ directions have the
expressions\cite{Barbier_2010},
\begin{eqnarray}
v_x ^j & = &\frac { j ^2 \pi ^2 } {\tilde{V} ^2}  v _ F \nonumber \\
v_y ^j & = & v_F - v_x ^j \, \, . \label{velocities}
\end{eqnarray}
\vspace{-0.5cm}
\subsubsection {Electrical conductivity.}  \vspace{-0.3cm}
The conductivity in the collisionless limit has the
expression \cite{Ziegler_2007,Ryu_2007}
\begin{equation}
\sigma _{\mu \mu} (\omega)\! \! = \! \! - i \frac {e^2}{\hbar} g_s
g_v \!\!\!\sum _{\mathbf{k},s,s'} { \frac {f _{\mathbf{k},s'} \!- \!
f_{\mathbf{k},s}}{\varepsilon _{\mathbf{k},s'} \! - \! \varepsilon
_{\mathbf{k},s}} \, \,   \frac {|<s,k|v _{\mu}|s',k>|^2
}{\varepsilon _{\mathbf{k},s'}\! - \! \varepsilon _{\mathbf{k},s}
\!-  \! \hbar \omega \! -  \!i \delta}} \label{sigma}
\end{equation}
where $s'$ and $s$ are band indices, $f _{\mathbf{k},s}$ is the
Fermi distribution function for the states $|s,\mathbf{k}>$,
$v_{\mu}$ is the velocity operator in the $\hat{\mu}$ direction and
$\delta$ is a positive infinitesimal constant. The conductivity
contains a factor $g_sg_v$=4, which takes into account the spin
and valley degeneracy. In the case of a single Dirac point with
anisotropic velocities $v_x$ and $v_y$,
expressed with a Dirac Hamiltonian of the form
\begin{equation}
H_A=\hbar(v_xk_x\sigma_x+v_yk_y\sigma_y),
\nonumber
\end{equation}
one may show that
the conductivity parallel and perpendicular to the
potential barriers of the superlattice
may be written in the form
\begin{eqnarray}
\sigma _{\parallel} ^0  ( \omega =0) & = &\frac {v_y ^0}{v_x ^0} \sigma _0
=
\sigma _0  \frac {| \sin (\tilde{V})|} {\tilde{V}}, \nonumber \\
\sigma _{\perp} ^0 (\omega =0) & = & \frac {v_x ^0}{v_y ^0} \sigma
_0=\sigma _0  \frac  {\tilde{V}} {|\sin (\tilde{V})|},
\label{sigma_0}
\end{eqnarray}
with $ \sigma _0$ the conductivity of an {\it isotropic} Dirac Hamiltonian.
The value of $\sigma _0$ depends on the order in which the zero frequency,
zero temperature and vanishing ``smearing parameter'' $\delta$ \cite{Ryu_2007} limits are
taken\cite{Ludwig_1994,Ryu_2007}.
However the form of the velocity rescaling of the
conductivity is independent of the order in which the limits are
taken.

In the case of several Dirac points in the spectrum, we assume that
each of the points contributes to the conductivity in parallel and
using Eq. (\ref{velocities}), the conductivity takes the form,
\begin{eqnarray}
\sigma _{\parallel} & = & \sigma _0 \left ( \frac {|\sin (\tilde{V})|}
{\tilde{V}} + 2 \sum _{j=1}^{j_{max}} { \frac {\tilde{V}^2-( \pi j)
^2}{( \pi
j ) ^2}} \right ) \nonumber \\
\sigma _{\perp} & = & \sigma _0 \left ( \frac  {\tilde{V}} {|\sin
(\tilde{V})|}  + 2 \sum _{j=1}^{j_{max}} { \frac {( \pi j ) ^2}
{\tilde{V}^2-( \pi j) ^2} } \right ) \label{conductivities}
\end{eqnarray}
where $j_{max}$=$\mathrm{Integer}(\frac {\tilde{V}}{ \pi})$ indicates the
number of Dirac point pairs induced by the superlattice. From this
expression we see that for small potentials the conductivity
perpendicular to the superlattice barriers increases quadratically with
$V_0d$, and each time a new pair of Dirac points emerges the
conductivity exhibits a peak. In the direction parallel to the
barriers, the conductivity decreases quadratically with
$V_0d$ and dips when new Dirac points emerge.

We remark that in obtaining Eq. (\ref{conductivities}), we
have assumed that each Dirac point contributes as an independent
channel to the conductivity and that near each Dirac point the
dispersion relation is linear over a wide range of the reciprocal
space.

\vspace{-0.3cm}
\subsubsection {Mode dependent transmission.}  \vspace{-0.3cm}
The conductivity of a system governed by the Dirac equation with
anisotropic velocities, $H=\hbar (v_x k_x \sigma _x + v _y k_y
\sigma _y)$, can be also obtained by calculating the transmission
probability of modes confined in a stripe of width $W$ and length
$L$ connected to heavily doped contacts\cite{Tworzydlo_2006,Katsnelson_2006a,Das_Sarma_2010}. For
transport along the $\hat{x}$-direction, the transmission
probability for a transverse mode has the form
\begin{equation}
T_n(\hat{x})=\frac 1 {\cosh ^2 ( \frac {v_y}{v_x} q_n L)} \, ,
\label{Trans_Prob}
\end{equation}
where the transverse momentum $q_n$ depends on the details of the
precise boundary condition of the strip 
\cite{Brey_2006b,Tworzydlo_2006}. For wide enough strips the
conductivity of the system is independent of the boundary
conditions and is found by summing over the modes,
\begin{eqnarray}
\sigma _{xx} & = & g_s g_v \frac L W \frac {e ^2}h \sum
_{n}T_n(\hat{x})= \frac {e^2} h \frac {2L}{\pi} \int _{-\infty}
^{\infty} \frac {dq} {\cosh ^2 ( \frac {v_y}{v_x} q L)} \nonumber
\\ & = & \frac 4 {\pi} \frac {e^2} h \frac {v_x}{v_y} \, \, \, \, \mathrm{for} \, \,\, W>>L \, \, . \label{int_trans}
\end{eqnarray}
The conductivity in the $\hat{y}$ direction is obtained by interchanging
$x$ and $y$ in the last equation. The condition for the
existence of a well defined -size independent- conductivity is the
dependence of the transmission probability on the product $qL$
(Eq. (\ref{Trans_Prob})) and the linear dispersion of the carriers. The
condition $W \gg L$ allows the sum the transmissions over the modes to
be written as
an integral over $q$ in Eq. (\ref{int_trans}).
\vspace{-0.3cm}
\subsubsection {Two dimensional superlattice potential.}  \vspace{-0.3cm}
We consider a two dimensional superlattice potential on top a
graphene sheet [as in Fig. \ref{fig:esquema}(c)]. In second order perturbation theory the group velocity of quasiparticles with momentum $\mathbf{k}$ has the
form \cite{Park_2008a}
\begin{equation}
v_ \mathbf{k} = v_F - v_F \sum _{\mathbf{G} \neq 0} \frac {2
|U(\mathbf{G})|^2 } {  \hbar ^2 v_F ^2 |\mathbf{G} | ^2} \sin ^2
\theta _{\mathbf{k},\mathbf{G}} \, \, \, , \label{v_SL_2D}
\end{equation}
where $\mathbf{G}$ and $U(\mathbf{G})$ are the reciprocal lattice
vectors and the corresponding Fourier component of the external
potential and $\theta _{\mathbf{k},\mathbf{G}}$ is the angle between
$\mathbf{G}$ and $\mathbf{k}$. Using the same approximation as in the
previous subsection the conductivity in the
$\hat{x}$-direction takes the form
\begin{equation}
\sigma_{xx} = \sigma _0 \frac  {\hbar ^2 v_F ^2 -\sum _{\mathbf{G}
\neq 0} 2 |U(\mathbf{G})|^2   \frac {G_y ^2} {   |\mathbf{G} | ^4} }
{\hbar ^2 v_F ^2 -\sum _{\mathbf{G} \neq 0} 2 |U(\mathbf{G})|^2
\frac {G_x ^2} {   |\mathbf{G} | ^4} } \, \, \, \, .
\label{sigmaxx_2D}
\end{equation}
The conductivity in the $\hat{y}$-direction is obtained  by
interchanging $G_x$ and $G_y$ in this expression. The
striking result of Eq. (\ref{sigmaxx_2D}) is that for
symmetric superlattice potentials the conductivity in the $\hat{x}$ and $\hat{y}$
directions are equal and take the value of pristine graphene, $\sigma
_{xx}$=$\sigma _{yy}$=$\sigma _0$=$\frac 4 {\pi} \frac { e^2} h  $.
The expression Eq. (\ref{v_SL_2D}) has been obtained in second order perturbation theory
and it is a good approximation provided that the superlattice potential does not induce new Dirac points.
We expect that Eq. (\ref{sigmaxx_2D}) will be valid in the same regime.

\section{Numerical calculations.}
\begin{figure}
\includegraphics[width=6.5cm]{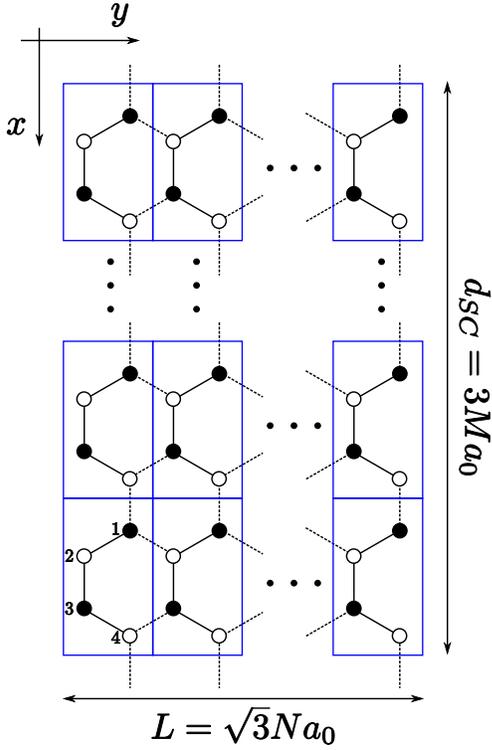}
\caption{(Color online) Schematic representation of the superlattice used in the tight-binding calculations showing the axis selection.
The system is infinite along the $x$-direction and has a finite length $L=\sqrt{3}Na_0$ along the $y$-direction. The superlattice has a vertical period $d=3M a_0$ in which periodic boundary conditions are imposed.}
\label{fig:armchair}
\end{figure}

In order to compute numerically the transport properties  we describe
the electronic states of a defect free graphene
layer using the tight-binding approximation,
\begin{equation}
\hat{H}
= - t_g \sum_{<ij>} \hat{c}^{\dagger}_{i} \hat{c}_{j}
+ \sum_{i} V_{i} \hat{c}^{\dagger}_{i}\hat{c}_{i}
\, \, ,
\end{equation}
where $t_g=2\hbar v_F/3a_0$ denotes the hopping element between nearest carbon
atoms on the hexagonal lattice, $a_0$ is the smallest carbon-carbon distance  and $V_i$ is the potential applied to the lattice.
The spin degree of freedom has been omitted due to degeneracy.

In order to analyze the different transport situations depicted in Fig. \ref{fig:esquema} we assume that the central region is a nanoribbon with armchair edges along the $\hat{x}$-direction as depicted in Fig. \ref{fig:armchair}. The nanoribbon is constructed by repeating a unit cell composed of four atoms $N$ times along the $\hat{y}$-direction and $M$ times along the $\hat{x}$-direction. Thus, the length of the graphene stripe is $L=N\sqrt{3}a_{0}$. For describing the $W\gg L$ limit we impose periodic boundary conditions in the transversal direction $\hat{x}$ and define $q\in  \left [-\pi/d_{SC},\pi/d_{SC} \right]$ as the corresponding wave vector, with $d_{SC}=3Ma_0$ being the vertical length of the supercell.

We connect the armchair edges of the nanoribbon to heavily doped graphene leads thus maintaining the graphene sublattice structure at the edges\cite{Schomerus_2007,Blanter_2007,Brey_2007c,Burset_2008,Burset_2009}.
The corresponding self-energies on the graphene sites at the layer
edges are approximated by a $4M\times 4M$ matrix with elements
$\Gamma^{L,R}_{ij,\alpha \beta}=\delta_{ij}\gamma^{L,R}_{\alpha
\beta}$, where $\alpha,\beta=1,\dots,4$ label the atomic sites
within the unit cell and $i,j=1,\dots,M$ label the unit cells in the
superlattice. Following the geometry depicted in Fig. \ref{fig:armchair}, the elements of the self-energy matrix are explicitly defined as
$\gamma^{L}_{22} = \gamma^{L}_{33} = \gamma^{R}_{11} = \gamma^{R}_{44} = i\sqrt{3}/2$ and $\gamma^{L}_{23} = \gamma^{L}_{32} = \gamma^{R}_{14} = \gamma^{R}_{41} = -1/2$\cite{Burset_2008}. Thus, we calculate the transmission at zero energy, $T(q)$, as
\begin{equation}
T(q) = 4 \mbox{Tr} \left[ \hat{\Gamma}_L \hat{G}^r_{LR} (E=0,q) \hat{\Gamma}_R \hat{G}^a_{RL} (E=0,q) \right],
\label{eq:trans}
\end{equation}
where $\hat{G}^{r,a}_{LR,RL} (E,q)$ are the $4M\times 4M$ retarded and advanced Green functions between the edges of the layer. Furthermore, for analyzing the transmission distribution it is useful to determine the eigenvalues $\tau_{\alpha}(q)$ of the transmission matrix $\hat{t}^{\dagger} \hat{t}$, where $\hat{t} = 2 \sqrt{\hat{\Gamma}_L} \hat{G}^r_{LR} (E,q) \sqrt{\hat{\Gamma}_R}$. From these eigenvalues one can determine the probability distribution $P(\tau) = \sum_{\alpha,q} \delta(\tau - \tau_{\alpha}(q))$ and the Fano factor
\begin{equation}
F=\frac{\sum\limits_{\alpha=1}^{4M} \sum\limits_{q} \tau_{\alpha}(q) \left( 1- \tau_{\alpha}(q) \right)}{\sum\limits_{\alpha=1}^{4M} \sum\limits_{q} \tau_{\alpha}(q)} .
\label{eq:fano}
\end{equation}
By integrating the transmission we compute the conductance of the system $G = (4e^2/\hbar) \int dq \mbox{Tr} \left[ \hat{t} \hat{t}^{\dagger} \right]$, where both the spin and valley degeneracies have been taken into account. The resulting conductivity, within the limit $W \gg L$, is obtained by multiplying by the geometrical factor $L$.

\subsection{Transport parallel to the superlattice barriers.}
For studying the transport parallel to the superlattice, we consider a periodic one-dimensional potential along the $\hat{x}$-direction within the previous geometry as is schematically depicted in Fig. \ref{fig:esquema}(a). The one-dimensional superlattice potential, $V_i$, has the piecewise constant form,
\begin{equation}
V_i = \left\{ \begin{array}{ccl}
V_{0} & , & 0 \leq x_i \leq \frac{d}{2} \\
-V_{0} & , & \frac{d}{2} < x_i \leq d
\end{array} \right. ,
\label{eq:potential}
\end{equation}
where $d=d_{SC}=3Ma_0$ is the period of the potential.

\begin{figure}
\includegraphics[width=8.5cm]{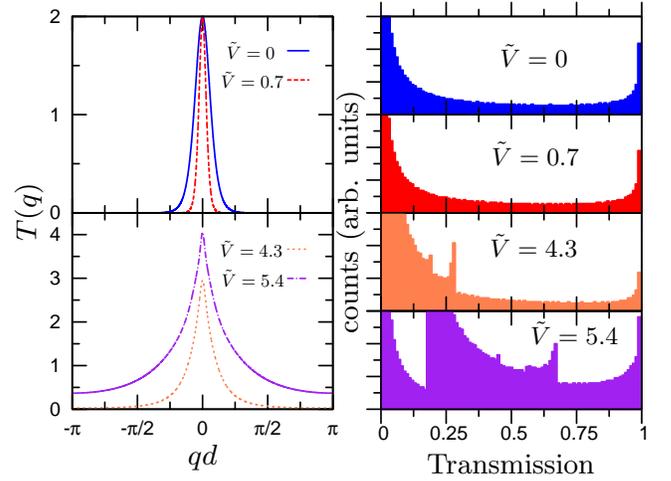}
\caption{(Color online)
In the left panels we plot, as function of $q d$, the transmission 
$T(q)$ per spin channel for a superlattice of period $d=42a_0$ and
amplitudes $\tilde{V}=\frac{V_0 d}{2\hbar v_F}=0$ and $\tilde{V}=0.7$ 
(top left panel) and for a period $d=54a_0$ and amplitudes $\tilde{V}=4.3$ and $\tilde{V}=5.4$ (bottom left panel). In the right panels we plot the distribution of the eigenvalues of the transmission matrix for the different values of $\tilde{V}$. The length of the stripe is $L=100\sqrt{3}a_0$.}
\label{fig:hist_hd1}
\end{figure}

In Fig. \ref{fig:hist_hd1} we plot the transmission $T(q)$ as function
of the product $q d$ for a superlattice of period $d=42 a_0$ and amplitudes  $V_0 d=0$ and $ V_0 d= 1.4\hbar v_F $ in the top left panel and for a period $d=54a_0$ and amplitudes $ V_0 d= 8.6 \hbar v_F $ and $ V_0 d= 10.8\hbar v_F $ in the bottom left panel. The horizontal length of the graphene layer is $L=100\sqrt{3}a_0$. We also plot in the right panels of Fig. \ref{fig:hist_hd1}  the distribution of the eigenvalues of the transmission matrix.

\begin{figure}[ht]
\includegraphics[width=8.5cm]{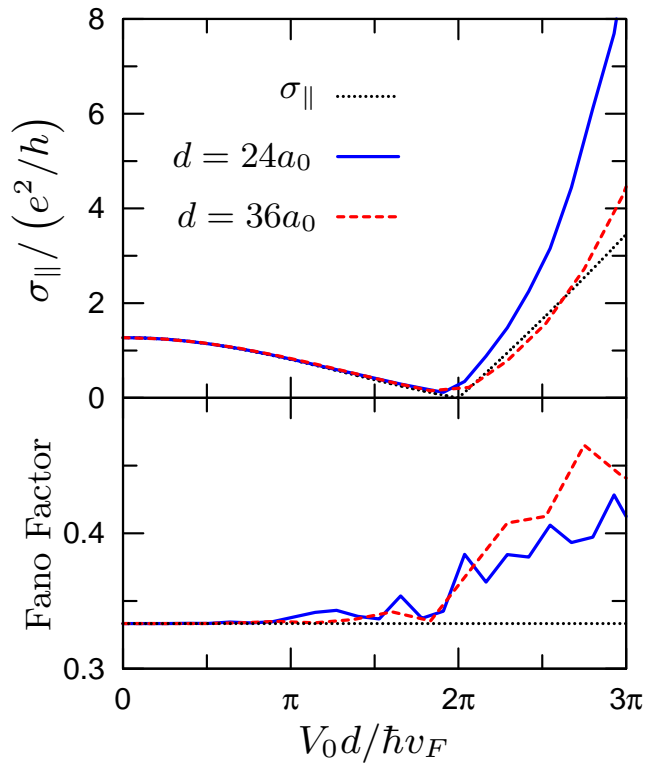}
\caption{(Color online) Transport parallel to the superlattice barriers. Top (bottom) panel shows the conductivity (Fano factor)  for a graphene sheet with $L=500\sqrt{3}a_0$ and superlattice period $d=24a_0$ (solid blue line) and $L=200\sqrt{3}a_0$ and superlattice period $d=36a_0$ (dashed red line) as a function of the normalized barrier height $V_0d$. Dotted line corresponds to the conductivity obtained in the continuum model assuming independent transport channels, Eq. (\ref{conductivities}), in the top panel and to the pseudo-diffusive value $F=1/3$ in the bottom panel.}
\label{fig:integral_y}
\end{figure}

In Fig. \ref{fig:integral_y} we plot, as function of $V_0 d$,
the conductivity  and the Fano factor obtained for a system of length $L$=$100 \sqrt{3} a_0$ and for different values of the superlattice period, $d$.

We first discuss the case of potential barriers in the range $V_0 < V_c =\frac {2 \pi \hbar v_F} {d}$ [top panels of Fig. \ref{fig:hist_hd1}].
For these superlattices the original Dirac points are the only active transport channels.
As a function of $q$ the transmission is peaked at $q$=0, and the width of the peak diminishes when $V_0 d$ increases.
The transmission fits very well to the functional form [see
Eq. (\ref{Trans_Prob})] $T(q) = 2/ \cosh^2 ( \frac   { \tilde{V} } {| \sin { \tilde{V}} | }qL)$, where the factor 2 accounts  for the valley degeneracy and $\tilde{V}=\frac {V_0 d }{ 2 \hbar v_F}$. The corresponding distribution of the eigenvalues of the transmission matrix has the form $P(\tau) \sim 1/\tau\sqrt{1-\tau}$ indicating the pseudo-diffusive character of the transport in this range of potentials.
The conductivity obtained by integrating the transmission is well-defined
and, in this range of $V_0d$, has the form $\sigma _ \parallel  ^0 = \sigma _0 \frac {| \sin (\tilde{V}) |}{\tilde{V}}$ [see Fig. \ref{fig:integral_y}]. The Fano factor in this range of potentials is 1/3 in agreement with the pseudo-diffusive character of transport.
We thus conclude that in the range of parameters $V_0 d < {2 \pi \hbar v_F} $ the transport is pseudo-diffusive, the conductivity only depends on the product $V_0 d$ and has
the form $\sigma _ \parallel ^0 = \sigma _0 \frac {| \sin (\tilde{V}) |}{\tilde{V}}$.

For normalized barrier heights $V_0d$ larger than $2 \pi \hbar v_F$
two new Dirac points per valley appear\cite{Brey_2009,Park_2009}. These new
Dirac points are new transmission channels in the system, that for
transport parallel to the superlattice barriers are superimposed in
reciprocal space upon the original Dirac points. The resultant
transmission exhibits a wider distribution in reciprocal space [see
bottom left panel of Fig. \ref{fig:hist_hd1}]. The width of the
transmission can reach the edges of the reduced Brillouin zone $\pm
\pi /d$ for small values of $L/d$. The corresponding distribution of
the eigenvalues of the transmission matrix is a superposition of the
distribution of each mode, and the corresponding Fano factor is
different than $1/3$. The conductivity should be independent on the
system size. We find that the value of $L$ where the conductivity is
well defined depends on $d$ and coincides with the value of $L$ in
which the transmission is non zero at the edges of the reduced
Brillouin zone. In Fig. \ref{fig:integral_y} we see that the general
trend of the conductivity for values of $V_0 d$ larger than $2 \pi
\hbar v_F$ is qualitatively described by the continuum model, Eq.
(\ref{conductivities}). However the analytical model neglects some
effects such as the coupling between the modes or the deviation from
linear dispersion, so that in this range of superlattice parameters
the conductivity depends separately on $V_0$ and $d$. The coupling
between the modes also leads to a Fano factor with a value larger
than $1/3$, and the transport is not pseudo-diffusive.

\subsection{Transport perpendicular to the superlattice barriers.}
In this section we consider a potential in the $\hat{y}$-direction and study the transport in the same direction, i.e. perpendicular to the
superlattice barriers [see Fig. \ref{fig:esquema}(b)]. Following the same geometry as in the previous section (see details in Fig. \ref{fig:armchair}), we define a one-dimensional piecewise potential along the $\hat{y}$-direction as
\begin{equation}
V_i = \left\{ \begin{array}{ccl}
V_{0} & , & 0 \leq y_i \leq \frac{d}{2} \\
-V_{0} & , & \frac{d}{2} < y_i \leq d
\end{array} \right. ,
\label{eq:potential_x}
\end{equation}
where $d=2n \sqrt{3} a_0$ is the period of the potential.

\begin{figure}[ht]
\includegraphics[width=8.5cm]{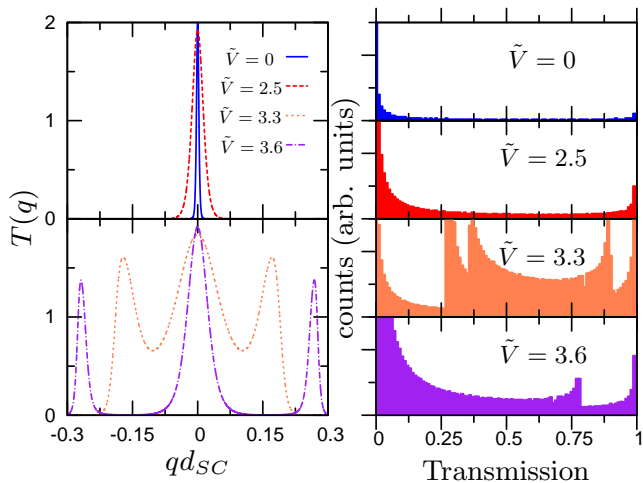}
\caption{(Color online)
In the left panels we plot, as function of $q d_{SC}$, the transmission  $T(q)$ per spin channel for a superlattice of period $d=38.1a_0$ and normalized amplitudes $\tilde{V}=\frac{V_0 d}{2\hbar v_F}=0$, $\tilde{V}=2.5$ (top left panel), $\tilde{V}=3.3$ and $\tilde{V}=3.6$ (bottom left panel).
In the right panels we plot the distribution of the eigenvalues of the transmission matrix for the different values of $\tilde{V}$.
The length of the stripe is $L=500\sqrt{3}a_0$.}
\label{fig:hist_hd2}
\end{figure}

In the left column of Fig. \ref{fig:hist_hd2} we plot the transmission $T(q)$ as a function of $q d_{SC}$ for a superlattice with period $d=38.1a_0$ and amplitudes $V_0 d=0$, $V_0 d=5\hbar v_F$ (top left panel), $V_0 d=6.6\hbar v_F$ and $V_0 d=7.2\hbar v_F$ (bottom left panel). The horizontal length of the graphene strip is $L=500\sqrt{3}a_0$. In the right column of Fig. \ref{fig:hist_hd2} we plot the corresponding distribution of the eigenvalues of the transmission matrix.

\begin{figure}[ht]
\includegraphics[width=8.5cm]{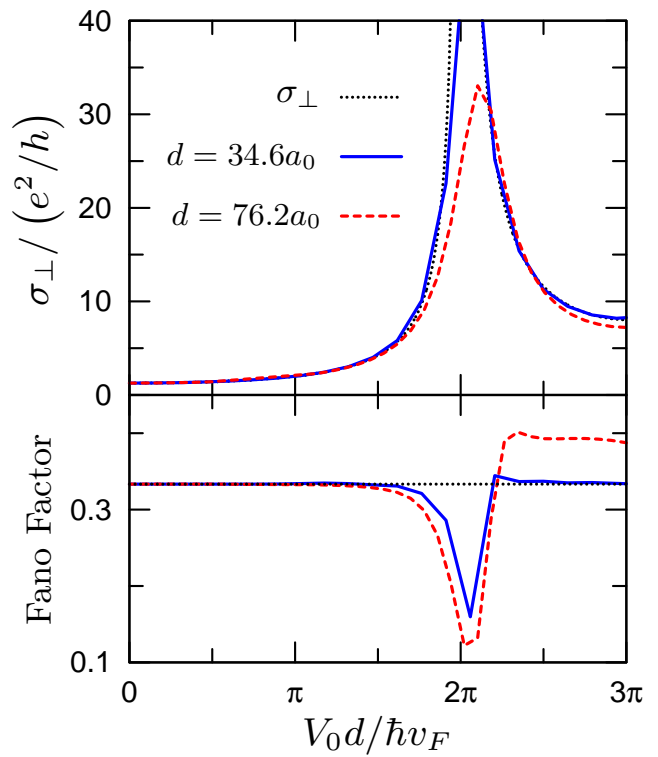}
\caption{(Color online) Transport perpendicular to the superlattice barriers. Top (bottom) panel shows the conductivity (Fano factor)  for a graphene sheet with $L=500\sqrt{3}a_0$ and superlattice periods $d=34.6a_0$ (solid blue line) $d=74.2a_0$ (dashed red line) as a function of the normalized barrier height $V_0d$. Dotted line corresponds to the conductivity obtained in the continuous model assuming independent transport channels, Eq. (\ref{conductivities}), in the top panel and to the pseudo-diffusive value $F=1/3$ in the bottom panel.}
\label{fig:integral_x}
\end{figure}

In the top panel of Fig. \ref{fig:integral_x} we show, as function of $V_0 d$,  the conductivity for horizontal periods of $d=34.6a_0$ and $d=76.2a_0$, for a graphene sheet of length $L$=$500\sqrt{3} a_0$. In the bottom panel of Fig. \ref{fig:integral_x} we plot the Fano factor for the same two values of the period of the superlattice.

In the range of potential barriers before the creation of new Dirac points, i.e. $V_0 < V_c$, the behavior of the transmission is exactly the inverse of the previous case. The contribution to the transmission from each valley is superimposed as a sharp peak at $q=0$. However, contrary to the previous result, the width of the peak increases with the product $V_0 d$. Following Eq. (\ref{Trans_Prob}), the transmission is fitted to $T(q) = 2/\cosh^2( \frac {| \sin{\tilde{V}} |} {\tilde{V}} qL)$. Subsequently, the distribution of the eigenvalues is that of pseudo-diffusive transport.
On the other hand, when $V_0 > 2\pi\hbar v_F$, a pair of Dirac points is created for each valley. In the bottom panel of Fig. \ref{fig:hist_hd2} we show how these new peaks split from the original ones until there are three almost independent contributions to the transmission. In this later case the distribution of eigenvalues for each mode returns to a form of the type $P(\tau) \sim 1/\tau\sqrt{1-\tau}$, indicative of pseudo-diffusive behavior. Before the new Dirac points are completely separated from the original ones, the coupling between modes produces a deviation from the pseudo-diffusive transport.

The behavior of the conductivity perpendicular to the barriers is completely different
than the parallel case. The perpendicular conductivity presents peaks at the values of the normalized potential height where new Dirac points appear. The numerical calculated conductivity agrees very well with the analytical one, Eq. (\ref{conductivities}), even for values of $V_0d > 2 \pi \hbar v_F$. The Fano factor has the value $1/3$ for all values of $V_0 d$ except near the values of $V_0d$ for which new Dirac appears. This indicates that, in this geometry, the Dirac points are weakly coupled and the approach of Section II for the conductivity is appropriate.

\subsection{Transport in a two dimensional superlattice}
\begin{figure}[ht]
\includegraphics[width=8.5cm]{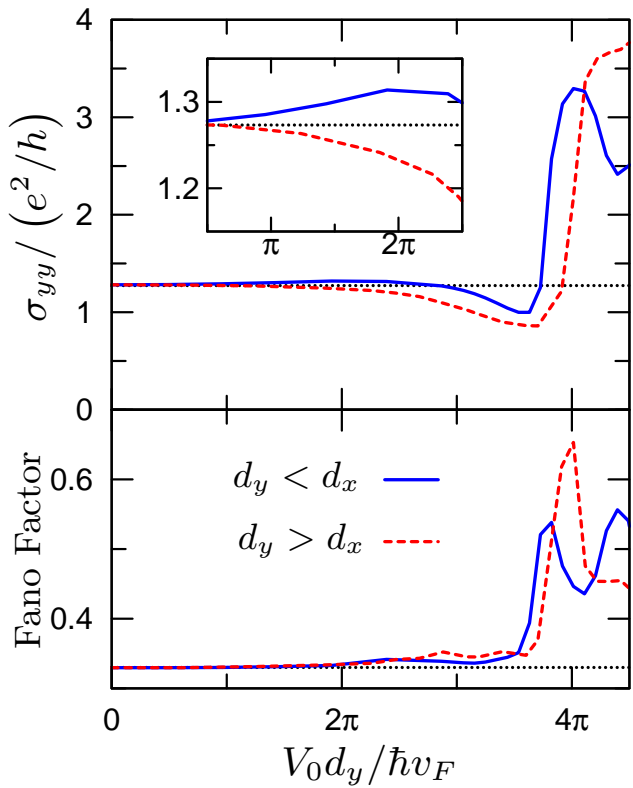}
\caption{(Color online) In the top (bottom) panel we plot the conductivity (Fano factor) as a function of the normalized barrier height $V_0d_y$ for a graphene sheet in presence of a two dimensional superlattice  with a fixed vertical period of $d_x=48a_{0}$ and different horizontal periods $d_y=45a_0$ (blue solid line) and $d_y=48.5a_0$ (red dashed line). The dotted line corresponds to the conductivity of pristine graphene ($\sigma_0$) in the top panel and to the pseudo-diffusive limit ($1/3$) in the bottom panel. The length of the graphene sheet is $L=200\sqrt{3}a_0$. Inset: the conductivity as a function of the normalized barrier height in the proximity of the critical potential $V_c$ in which a new pair of Dirac points is created.}
\label{fig:integral_xy}
\end{figure}

One of the more striking results presented in Section II is that the conductivity of graphene in the presence of a symmetric two dimensional
superlattice potential is $\frac 4 {\pi} \frac {e ^2} {\hbar}$ independent of the period and the height of the potential barriers. In order to check this result we have built a chessboard-like potential combining piecewise potentials in the $\hat{x}$, Eq. (\ref{eq:potential}), and $\hat{y}$, Eq. (\ref{eq:potential_x}), directions in a way in which a potential barrier is always followed by a well along each direction (see Fig. \ref{fig:esquema}(c)). The length of the period in the $\hat{x}$ and $\hat{y}$ directions is $d_x$ and $d_y$ respectively. Because the underlying triangular lattice of graphene, the period in both directions cannot be exactly equal.

The top panel of Fig. \ref{fig:integral_xy} shows the conductivity as a function of the potential height $V_0$ for a graphene layer with $L=200\sqrt{3}a_0$ and a fixed vertical period of $d_x=48a_{0}$. We plot the conductivity for two different horizontal periods $d_y=45a_0$ and $d_y=48.5a_0$. We compare these results with the isotropic conductivity of graphene $\sigma_0=\frac{4}{\pi}\frac{e^2}{h}$.

A remarkable result is that the conductivity in this potential remains almost constant in the range $V_0 \gtrsim V_c$ where a new pair of Dirac points is created in the previously studied cases. Thus, in this range of potential barriers, Eq. (\ref{sigmaxx_2D}) obtained in second order perturbation theory remains a good approximation according to the tight-binding results. Furthermore, the pseudo-diffusive behavior of transport is maintained for a large range of the potential barriers. In the bottom panel of Fig. \ref{fig:integral_xy} we show how the Fano factor is stable around the pseudo-diffusive value of $1/3$ while $V_0d_y \lesssim 4\pi \hbar v_F$. When $V_0d_y \sim 4\pi \hbar v_F$, which for the previous potentials corresponded to the creation of the second pair of Dirac points, the conductivity deviates from $\sigma_0$, the Fano factor increases and transport is no longer pseudo-diffusive. The approximation of weakly coupled Dirac points is then no longer applicable.

The small deviations from the conductivity of pristine graphene that occurs when $V_0d_y \sim 2\pi \hbar v_F$ can be more clearly appreciated in the inset of Fig. \ref{fig:integral_xy}. Due to the geometry of the graphene layer, the period in both directions is never exactly the same. This affects the validity of Eq. (\ref{sigmaxx_2D}) to a small degree. When $d_y \lesssim d_x$ the conductivity slightly increases from $\sigma_0$, presenting a positive slope, while if $d_y \gtrsim d_x$ the effect is the opposite. When the difference between both periods becomes larger the conductivity continuously evolves into the corresponding case of the previous sections (Figs. \ref{fig:integral_y},\ref{fig:integral_x}).

\vspace{0.3cm}
\section{Conclusions}
\vspace{0.3cm}
Superlattice potentials generically induce anisotropy in the dispersions near the
Dirac points in graphene, and under certain circumstances may induce
extra Dirac points at zero energy.  In this work we demonstrated that
when the Fermi energy passes through a spectrum with a single
anisotropic Dirac point, the resulting conductivity can be expressed
in a very simple way in terms of the velocities along the two
principle directions of the anisotropy, and the conductivity for
the corresponding isotropic Dirac point.  The result can be generalized
to the case of several Dirac points when they are sufficiently
separated in momentum space so that a conductivity expressed
as a sum over those of independent Dirac points is sensible.
For a two-dimensional superlattice which induces little
anisotropy in the spectrum, a remarkable result is that
the conductivity is essentially unchanged from the result
for pristine graphene, even if the velocity renormalization
is quite large.

Numerical tight-binding calculations generally confirm this
simple picture.  In particular one finds the conductivity
parallel and perpendicular to the superlattice barriers
for a one-dimensional potential evolve in opposite directions
with increasing $V_0d$, and that for a spectrum in which
no new Dirac points have been generated there is quantitative
agreement with the simple analytical model.  As new Dirac points
are introduced into the spectrum one finds dips in $\sigma_{\parallel}$
and peaks in $\sigma_{\perp}$ as expected, although the results
are less quantitatively described by the continuum model, presumably
because the wavefunctions cannot be uniquely associated with
single Dirac points.  Deviations of the Fano factor from
pseudo-diffusive behavior confirm this interpretation.

These studies suggest that more complicated potentials could
also yield behaviors in the conductance with simple interpretations.
For example, a modulated superlattice potential yields a Landau
level spectrum \cite{Sun_2010}, for which $\sigma_{\parallel}$ may have behavior reminiscent
of edge state transport \cite{Iyengar_unpub}.  It is also interesting
to speculate that for isotropic superlattice potentials, one may
sufficiently slow the electron velocity so that electron-electron
interaction effects become important \cite{Jianhui_2010,Jianhui_2011}.
We leave these questions for future research.

\section*{Acknowledgments}
PB, LB and ALY thank Juanjo Palacios for helpful discussions.
Funding for the work described here was provided by MICINN-Spain via
grants FIS2009-08744 (LB) and FIS2008-04209 (PB and ALY), and by the NSF
through Grant No. DMR-1005035 (HAF).


\begin{thebibliography}{51}
\expandafter\ifx\csname natexlab\endcsname\relax\def\natexlab#1{#1}\fi
\expandafter\ifx\csname bibnamefont\endcsname\relax
  \def\bibnamefont#1{#1}\fi
\expandafter\ifx\csname bibfnamefont\endcsname\relax
  \def\bibfnamefont#1{#1}\fi
\expandafter\ifx\csname citenamefont\endcsname\relax
  \def\citenamefont#1{#1}\fi
\expandafter\ifx\csname url\endcsname\relax
  \def\url#1{\texttt{#1}}\fi
\expandafter\ifx\csname urlprefix\endcsname\relax\def\urlprefix{URL }\fi
\providecommand{\bibinfo}[2]{#2}
\providecommand{\eprint}[2][]{\url{#2}}

\bibitem[{\citenamefont{Geim and Novoselov}(2007)}]{Geim_2007}
\bibinfo{author}{\bibfnamefont{A.~K.} \bibnamefont{Geim}} \bibnamefont{and}
  \bibinfo{author}{\bibfnamefont{K.~S.} \bibnamefont{Novoselov}},
  \bibinfo{journal}{Nat.Mat.} \textbf{\bibinfo{volume}{6}},
  \bibinfo{pages}{183} (\bibinfo{year}{2007}).

\bibitem[{\citenamefont{Castro-Neto et~al.}(2009)\citenamefont{Castro-Neto,
  Guinea, Peres, Novoselov, and Geim}}]{Castro_Neto_RMP}
\bibinfo{author}{\bibfnamefont{A.~H.} \bibnamefont{Castro-Neto}},
  \bibinfo{author}{\bibfnamefont{F.}~\bibnamefont{Guinea}},
  \bibinfo{author}{\bibfnamefont{N.~M.~R.} \bibnamefont{Peres}},
  \bibinfo{author}{\bibfnamefont{K.~S.} \bibnamefont{Novoselov}},
  \bibnamefont{and} \bibinfo{author}{\bibfnamefont{A.~K.} \bibnamefont{Geim}},
  \bibinfo{journal}{Rev.\ Mod.\ Phys.} \textbf{\bibinfo{volume}{81}},
  \bibinfo{pages}{109} (\bibinfo{year}{2009}).

\bibitem[{\citenamefont{Novoselov et~al.}(2005)\citenamefont{Novoselov, Jiang,
  Booth, Khotkevich, Morozov, and Geim}}]{Novoselov_2005}
\bibinfo{author}{\bibfnamefont{K.~S.} \bibnamefont{Novoselov}},
  \bibinfo{author}{\bibfnamefont{D.}~\bibnamefont{Jiang}},
  \bibinfo{author}{\bibfnamefont{T.}~\bibnamefont{Booth}},
  \bibinfo{author}{\bibfnamefont{V.~V.} \bibnamefont{Khotkevich}},
  \bibinfo{author}{\bibfnamefont{S.~M.} \bibnamefont{Morozov}},
  \bibnamefont{and} \bibinfo{author}{\bibfnamefont{A.~K.} \bibnamefont{Geim}},
  \bibinfo{journal}{Nature} \textbf{\bibinfo{volume}{438}},
  \bibinfo{pages}{197} (\bibinfo{year}{2005}).

\bibitem[{\citenamefont{Zhang et~al.}(2005)\citenamefont{Zhang, Tan, Stormer,
  and Kim}}]{Zhang_2005}
\bibinfo{author}{\bibfnamefont{Y.}~\bibnamefont{Zhang}},
  \bibinfo{author}{\bibfnamefont{Y.-W.} \bibnamefont{Tan}},
  \bibinfo{author}{\bibfnamefont{H.~L.} \bibnamefont{Stormer}},
  \bibnamefont{and} \bibinfo{author}{\bibfnamefont{P.}~\bibnamefont{Kim}},
  \bibinfo{journal}{Nature} \textbf{\bibinfo{volume}{438}},
  \bibinfo{pages}{201} (\bibinfo{year}{2005}).

\bibitem[{\citenamefont{Kastnelson et~al.}(2006)\citenamefont{Kastnelson,
  Novoselov, and A.K.Geim}}]{Kastnelson_2006}
\bibinfo{author}{\bibfnamefont{M.~I.} \bibnamefont{Kastnelson}},
  \bibinfo{author}{\bibfnamefont{K.~S.} \bibnamefont{Novoselov}},
  \bibnamefont{and} \bibinfo{author}{\bibnamefont{A.K.Geim}},
  \bibinfo{journal}{Nat.Phys.} \textbf{\bibinfo{volume}{2}},
  \bibinfo{pages}{620} (\bibinfo{year}{2006}).

\bibitem[{\citenamefont{Huard et~al.}(2007)\citenamefont{Huard, Sulpizio,
  Stander, Todd, Yang, and Goldhaber-Gordon}}]{Huard_2007}
\bibinfo{author}{\bibfnamefont{B.}~\bibnamefont{Huard}},
  \bibinfo{author}{\bibfnamefont{J.~A.} \bibnamefont{Sulpizio}},
  \bibinfo{author}{\bibfnamefont{N.}~\bibnamefont{Stander}},
  \bibinfo{author}{\bibfnamefont{K.}~\bibnamefont{Todd}},
  \bibinfo{author}{\bibfnamefont{B.}~\bibnamefont{Yang}}, \bibnamefont{and}
  \bibinfo{author}{\bibfnamefont{D.}~\bibnamefont{Goldhaber-Gordon}},
  \bibinfo{journal}{Phys. Rev. Lett.} \textbf{\bibinfo{volume}{98}},
  \bibinfo{pages}{236803} (\bibinfo{year}{2007}).

\bibitem[{\citenamefont{Young and Kim}(2009)}]{Young_2009}
\bibinfo{author}{\bibfnamefont{A.~F.} \bibnamefont{Young}} \bibnamefont{and}
  \bibinfo{author}{\bibfnamefont{P.}~\bibnamefont{Kim}},
  \bibinfo{journal}{Nature Phys.} \textbf{\bibinfo{volume}{5}},
  \bibinfo{pages}{222} (\bibinfo{year}{2009}).

\bibitem[{\citenamefont{Stander et~al.}(2009)\citenamefont{Stander, Huard, and
  Goldhaber-Gordon}}]{Stander_2009}
\bibinfo{author}{\bibfnamefont{N.}~\bibnamefont{Stander}},
  \bibinfo{author}{\bibfnamefont{B.}~\bibnamefont{Huard}}, \bibnamefont{and}
  \bibinfo{author}{\bibfnamefont{D.}~\bibnamefont{Goldhaber-Gordon}},
  \bibinfo{journal}{Phys. Rev. Lett.} \textbf{\bibinfo{volume}{102}},
  \bibinfo{pages}{026807} (\bibinfo{year}{2009}).

\bibitem[{\citenamefont{Jr et~al.}(2009)\citenamefont{Jr, Liu, Bao, and
  Lau}}]{Velasco_2009}
\bibinfo{author}{\bibfnamefont{J.~V.} \bibnamefont{Jr}},
  \bibinfo{author}{\bibfnamefont{G.}~\bibnamefont{Liu}},
  \bibinfo{author}{\bibfnamefont{W.}~\bibnamefont{Bao}}, \bibnamefont{and}
  \bibinfo{author}{\bibfnamefont{C.~N.} \bibnamefont{Lau}},
  \bibinfo{journal}{New Journal of Physics} \textbf{\bibinfo{volume}{11}},
  \bibinfo{pages}{095008} (\bibinfo{year}{2009}).

\bibitem[{\citenamefont{Russo et~al.}(2009)\citenamefont{Russo, Craciun,
  Yamamoto, Tarucha, and Morpurgo}}]{Russo_2009}
\bibinfo{author}{\bibfnamefont{S.}~\bibnamefont{Russo}},
  \bibinfo{author}{\bibfnamefont{M.~F.} \bibnamefont{Craciun}},
  \bibinfo{author}{\bibfnamefont{M.}~\bibnamefont{Yamamoto}},
  \bibinfo{author}{\bibfnamefont{S.}~\bibnamefont{Tarucha}}, \bibnamefont{and}
  \bibinfo{author}{\bibfnamefont{A.~F.} \bibnamefont{Morpurgo}},
  \bibinfo{journal}{New Journal of Physics} \textbf{\bibinfo{volume}{11}},
  \bibinfo{pages}{095018} (\bibinfo{year}{2009}).

\bibitem[{\citenamefont{Pereira et~al.}(2006)\citenamefont{Pereira, Mlinar,
  Peeters, and Vasilopoulos}}]{Pereira_2006}
\bibinfo{author}{\bibfnamefont{J.~M.} \bibnamefont{Pereira}},
  \bibinfo{author}{\bibfnamefont{V.}~\bibnamefont{Mlinar}},
  \bibinfo{author}{\bibfnamefont{F.~M.} \bibnamefont{Peeters}},
  \bibnamefont{and}
  \bibinfo{author}{\bibfnamefont{P.}~\bibnamefont{Vasilopoulos}},
  \bibinfo{journal}{Phys. Rev. B} \textbf{\bibinfo{volume}{74}},
  \bibinfo{pages}{045424} (\bibinfo{year}{2006}).

\bibitem[{\citenamefont{Cheianov et~al.}(2007)\citenamefont{Cheianov, Fal'ko,
  and Altshuler}}]{Cheianov_2007}
\bibinfo{author}{\bibfnamefont{V.~V.} \bibnamefont{Cheianov}},
  \bibinfo{author}{\bibfnamefont{V.}~\bibnamefont{Fal'ko}}, \bibnamefont{and}
  \bibinfo{author}{\bibfnamefont{B.~L.} \bibnamefont{Altshuler}},
  \bibinfo{journal}{Science} \textbf{\bibinfo{volume}{315}},
  \bibinfo{pages}{1252} (\bibinfo{year}{2007}).

\bibitem[{\citenamefont{Zhang and Fogler}(2008)}]{Zhang_2008}
\bibinfo{author}{\bibfnamefont{L.~M.} \bibnamefont{Zhang}} \bibnamefont{and}
  \bibinfo{author}{\bibfnamefont{M.~M.} \bibnamefont{Fogler}},
  \bibinfo{journal}{Phys. Rev. Lett.} \textbf{\bibinfo{volume}{100}},
  \bibinfo{pages}{116804} (\bibinfo{year}{2008}).

\bibitem[{\citenamefont{Beenakker}(2008)}]{Beenakker_2008}
\bibinfo{author}{\bibfnamefont{C.~W.~J.} \bibnamefont{Beenakker}},
  \bibinfo{journal}{Rev. Mod. Phys.} \textbf{\bibinfo{volume}{80}},
  \bibinfo{pages}{1337} (\bibinfo{year}{2008}).

\bibitem[{\citenamefont{Arovas et~al.}(2010)\citenamefont{Arovas, Brey, Fertig,
  Kim, and Ziegler}}]{Arovas_2010}
\bibinfo{author}{\bibfnamefont{D.~P.} \bibnamefont{Arovas}},
  \bibinfo{author}{\bibfnamefont{L.}~\bibnamefont{Brey}},
  \bibinfo{author}{\bibfnamefont{H.~A.} \bibnamefont{Fertig}},
  \bibinfo{author}{\bibfnamefont{E.-A.} \bibnamefont{Kim}}, \bibnamefont{and}
  \bibinfo{author}{\bibfnamefont{K.}~\bibnamefont{Ziegler}},
  \bibinfo{journal}{New Journal of Physics} \textbf{\bibinfo{volume}{12}},
  \bibinfo{pages}{123020} (\bibinfo{year}{2010}).

\bibitem[{\citenamefont{V\'azquez~de Parga
  et~al.}(2008)\citenamefont{V\'azquez~de Parga, Calleja, Borca, Passeggi,
  Hinarejos, Guinea, and Miranda}}]{Vazquez_2008}
\bibinfo{author}{\bibfnamefont{A.~L.} \bibnamefont{V\'azquez~de Parga}},
  \bibinfo{author}{\bibfnamefont{F.}~\bibnamefont{Calleja}},
  \bibinfo{author}{\bibfnamefont{B.}~\bibnamefont{Borca}},
  \bibinfo{author}{\bibfnamefont{M.~C.~G.} \bibnamefont{Passeggi}},
  \bibinfo{author}{\bibfnamefont{J.~J.} \bibnamefont{Hinarejos}},
  \bibinfo{author}{\bibfnamefont{F.}~\bibnamefont{Guinea}}, \bibnamefont{and}
  \bibinfo{author}{\bibfnamefont{R.}~\bibnamefont{Miranda}},
  \bibinfo{journal}{Phys. Rev. Lett.} \textbf{\bibinfo{volume}{100}},
  \bibinfo{pages}{056807} (\bibinfo{year}{2008}).

\bibitem[{\citenamefont{Pletikosi\ifmmode~\acute{c}\else \'{c}\fi{}
  et~al.}(2009)\citenamefont{Pletikosi\ifmmode~\acute{c}\else \'{c}\fi{},
  Kralj, Pervan, Brako, Coraux, NDiaye, Busse, and Michely}}]{Pletikosic_2009}
\bibinfo{author}{\bibfnamefont{I.}~\bibnamefont{Pletikosi\ifmmode~\acute{c}\else
  \'{c}\fi{}}}, \bibinfo{author}{\bibfnamefont{M.}~\bibnamefont{Kralj}},
  \bibinfo{author}{\bibfnamefont{P.}~\bibnamefont{Pervan}},
  \bibinfo{author}{\bibfnamefont{R.}~\bibnamefont{Brako}},
  \bibinfo{author}{\bibfnamefont{J.}~\bibnamefont{Coraux}},
  \bibinfo{author}{\bibfnamefont{A.~T.} \bibnamefont{NDiaye}},
  \bibinfo{author}{\bibfnamefont{C.}~\bibnamefont{Busse}}, \bibnamefont{and}
  \bibinfo{author}{\bibfnamefont{T.}~\bibnamefont{Michely}},
  \bibinfo{journal}{Phys. Rev. Lett.} \textbf{\bibinfo{volume}{102}},
  \bibinfo{pages}{056808} (\bibinfo{year}{2009}).

\bibitem[{\citenamefont{{Tiwari} and {Stroud}}(2009)}]{Tivari_2009}
\bibinfo{author}{\bibfnamefont{R.~P.} \bibnamefont{{Tiwari}}} \bibnamefont{and}
  \bibinfo{author}{\bibfnamefont{D.}~\bibnamefont{{Stroud}}},
  \bibinfo{journal}{\prb} \textbf{\bibinfo{volume}{79}},
  \bibinfo{pages}{205435} (\bibinfo{year}{2009}).

\bibitem[{\citenamefont{Guinea and Low}(2010)}]{Guinea_2010}
\bibinfo{author}{\bibfnamefont{F.}~\bibnamefont{Guinea}} \bibnamefont{and}
  \bibinfo{author}{\bibfnamefont{T.}~\bibnamefont{Low}},
  \bibinfo{journal}{Philosophical Transactions of the Royal Society A:
  Mathematical, Physical and Engineering Sciences}
  \textbf{\bibinfo{volume}{368}}, \bibinfo{pages}{5391} (\bibinfo{year}{2010}).

\bibitem[{\citenamefont{Park et~al.}(2008{\natexlab{a}})\citenamefont{Park,
  Yang, Son, Cohen, and Louie}}]{Park_2008c}
\bibinfo{author}{\bibfnamefont{C.-H.} \bibnamefont{Park}},
  \bibinfo{author}{\bibfnamefont{L.}~\bibnamefont{Yang}},
  \bibinfo{author}{\bibfnamefont{Y.-W.} \bibnamefont{Son}},
  \bibinfo{author}{\bibfnamefont{M.~L.} \bibnamefont{Cohen}}, \bibnamefont{and}
  \bibinfo{author}{\bibfnamefont{S.~G.} \bibnamefont{Louie}},
  \bibinfo{journal}{Phys.Rev.Lett.} \textbf{\bibinfo{volume}{101}},
  \bibinfo{pages}{126804} (\bibinfo{year}{2008}{\natexlab{a}}).

\bibitem[{\citenamefont{Park et~al.}(2008{\natexlab{b}})\citenamefont{Park,
  Yang, Son, Cohen, and Louie}}]{Park_2008a}
\bibinfo{author}{\bibfnamefont{C.-H.} \bibnamefont{Park}},
  \bibinfo{author}{\bibfnamefont{L.}~\bibnamefont{Yang}},
  \bibinfo{author}{\bibfnamefont{Y.-W.} \bibnamefont{Son}},
  \bibinfo{author}{\bibfnamefont{M.~L.} \bibnamefont{Cohen}}, \bibnamefont{and}
  \bibinfo{author}{\bibfnamefont{S.~G.} \bibnamefont{Louie}},
  \bibinfo{journal}{Nat.Phys.} \textbf{\bibinfo{volume}{4}},
  \bibinfo{pages}{213} (\bibinfo{year}{2008}{\natexlab{b}}).

\bibitem[{\citenamefont{Park et~al.}(2008{\natexlab{c}})\citenamefont{Park,
  Son, Yang, Cohen, and Louie}}]{Park_2008b}
\bibinfo{author}{\bibfnamefont{C.-H.} \bibnamefont{Park}},
  \bibinfo{author}{\bibfnamefont{Y.-W.} \bibnamefont{Son}},
  \bibinfo{author}{\bibfnamefont{L.}~\bibnamefont{Yang}},
  \bibinfo{author}{\bibfnamefont{M.~L.} \bibnamefont{Cohen}}, \bibnamefont{and}
  \bibinfo{author}{\bibfnamefont{S.~G.} \bibnamefont{Louie}},
  \bibinfo{journal}{Nano Lett.} \textbf{\bibinfo{volume}{8}},
  \bibinfo{pages}{2020} (\bibinfo{year}{2008}{\natexlab{c}}).

\bibitem[{\citenamefont{Barbier et~al.}(2008)\citenamefont{Barbier, Peeters,
  Vasilopoulos, and Pereira}}]{Barbier_2008}
\bibinfo{author}{\bibfnamefont{M.}~\bibnamefont{Barbier}},
  \bibinfo{author}{\bibfnamefont{F.~M.} \bibnamefont{Peeters}},
  \bibinfo{author}{\bibfnamefont{P.}~\bibnamefont{Vasilopoulos}},
  \bibnamefont{and} \bibinfo{author}{\bibfnamefont{J.~M.}
  \bibnamefont{Pereira}}, \bibinfo{journal}{Phys.\ Rev.\ B}
  \textbf{\bibinfo{volume}{77}}, \bibinfo{pages}{115446}
  (\bibinfo{year}{2008}).

\bibitem[{\citenamefont{Brey and Fertig}(2009)}]{Brey_2009}
\bibinfo{author}{\bibfnamefont{L.}~\bibnamefont{Brey}} \bibnamefont{and}
  \bibinfo{author}{\bibfnamefont{H.~A.} \bibnamefont{Fertig}},
  \bibinfo{journal}{Phys. Rev. Lett.} \textbf{\bibinfo{volume}{103}},
  \bibinfo{pages}{046809} (\bibinfo{year}{2009}).

\bibitem[{\citenamefont{Park et~al.}(2009)\citenamefont{Park, Son, Yang, Cohen,
  and Louie}}]{Park_2009}
\bibinfo{author}{\bibfnamefont{C.-H.} \bibnamefont{Park}},
  \bibinfo{author}{\bibfnamefont{Y.-W.} \bibnamefont{Son}},
  \bibinfo{author}{\bibfnamefont{L.}~\bibnamefont{Yang}},
  \bibinfo{author}{\bibfnamefont{M.~L.} \bibnamefont{Cohen}}, \bibnamefont{and}
  \bibinfo{author}{\bibfnamefont{S.~G.} \bibnamefont{Louie}},
  \bibinfo{journal}{Phys. Rev. Lett.} \textbf{\bibinfo{volume}{103}},
  \bibinfo{pages}{046808} (\bibinfo{year}{2009}).

\bibitem[{\citenamefont{Barbier et~al.}(2010)\citenamefont{Barbier,
  Vasilopoulos, and Peeters}}]{Barbier_2010}
\bibinfo{author}{\bibfnamefont{M.}~\bibnamefont{Barbier}},
  \bibinfo{author}{\bibfnamefont{P.}~\bibnamefont{Vasilopoulos}},
  \bibnamefont{and} \bibinfo{author}{\bibfnamefont{F.~M.}
  \bibnamefont{Peeters}}, \bibinfo{journal}{Phys. Rev. B}
  \textbf{\bibinfo{volume}{81}}, \bibinfo{pages}{075438}
  (\bibinfo{year}{2010}).

\bibitem[{\citenamefont{Sun et~al.}(2010)\citenamefont{Sun, Fertig, and
  Brey}}]{Sun_2010}
\bibinfo{author}{\bibfnamefont{J.}~\bibnamefont{Sun}},
  \bibinfo{author}{\bibfnamefont{H.~A.} \bibnamefont{Fertig}},
  \bibnamefont{and} \bibinfo{author}{\bibfnamefont{L.}~\bibnamefont{Brey}},
  \bibinfo{journal}{Phys. Rev. Lett.} \textbf{\bibinfo{volume}{105}},
  \bibinfo{pages}{156801} (\bibinfo{year}{2010}).

\bibitem[{\citenamefont{Dell'Anna and De~Martino}(2009)}]{Martino_2009}
\bibinfo{author}{\bibfnamefont{L.}~\bibnamefont{Dell'Anna}} \bibnamefont{and}
  \bibinfo{author}{\bibfnamefont{A.}~\bibnamefont{De~Martino}},
  \bibinfo{journal}{Phys. Rev. B} \textbf{\bibinfo{volume}{79}},
  \bibinfo{pages}{045420} (\bibinfo{year}{2009}).

\bibitem[{\citenamefont{Snyman}(2009)}]{Snyman_2009}
\bibinfo{author}{\bibfnamefont{I.}~\bibnamefont{Snyman}},
  \bibinfo{journal}{Phys. Rev. B} \textbf{\bibinfo{volume}{80}},
  \bibinfo{pages}{054303} (\bibinfo{year}{2009}).

\bibitem[{\citenamefont{Tan et~al.}(2010)\citenamefont{Tan, Park, and
  Louie}}]{Tan_2010}
\bibinfo{author}{\bibfnamefont{L.~Z.} \bibnamefont{Tan}},
  \bibinfo{author}{\bibfnamefont{C.-H.} \bibnamefont{Park}}, \bibnamefont{and}
  \bibinfo{author}{\bibfnamefont{S.~G.} \bibnamefont{Louie}},
  \bibinfo{journal}{Phys. Rev. B} \textbf{\bibinfo{volume}{81}},
  \bibinfo{pages}{195426} (\bibinfo{year}{2010}).

\bibitem[{\citenamefont{Dell'Anna and De~Martino}()}]{Martino_2011}
\bibinfo{author}{\bibfnamefont{L.}~\bibnamefont{Dell'Anna}} \bibnamefont{and}
  \bibinfo{author}{\bibfnamefont{A.}~\bibnamefont{De~Martino}},
  \eprint{arXiv:1101.1918v1}.

\bibitem[{\citenamefont{Abedpour et~al.}(2009)\citenamefont{Abedpour,
  Esmailpour, Asgari, and Tabar}}]{Abedpour_2009}
\bibinfo{author}{\bibfnamefont{N.}~\bibnamefont{Abedpour}},
  \bibinfo{author}{\bibfnamefont{A.}~\bibnamefont{Esmailpour}},
  \bibinfo{author}{\bibfnamefont{R.}~\bibnamefont{Asgari}}, \bibnamefont{and}
  \bibinfo{author}{\bibfnamefont{M.~R.~R.} \bibnamefont{Tabar}},
  \bibinfo{journal}{Phys. Rev. B} \textbf{\bibinfo{volume}{79}},
  \bibinfo{pages}{165412} (\bibinfo{year}{2009}).

\bibitem[{\citenamefont{Barbier et~al.}(2009)\citenamefont{Barbier,
  Vasilopoulos, and Peeters}}]{Barbier_2009}
\bibinfo{author}{\bibfnamefont{M.}~\bibnamefont{Barbier}},
  \bibinfo{author}{\bibfnamefont{P.}~\bibnamefont{Vasilopoulos}},
  \bibnamefont{and} \bibinfo{author}{\bibfnamefont{F.~M.}
  \bibnamefont{Peeters}}, \bibinfo{journal}{Phys. Rev. B}
  \textbf{\bibinfo{volume}{80}}, \bibinfo{pages}{205415}
  (\bibinfo{year}{2009}).

\bibitem[{\citenamefont{Wang and Zhu}(2010)}]{Wang_2010}
\bibinfo{author}{\bibfnamefont{L.-G.} \bibnamefont{Wang}} \bibnamefont{and}
  \bibinfo{author}{\bibfnamefont{S.-Y.} \bibnamefont{Zhu}},
  \bibinfo{journal}{Phys. Rev. B} \textbf{\bibinfo{volume}{81}},
  \bibinfo{pages}{205444} (\bibinfo{year}{2010}).

\bibitem[{\citenamefont{Yampol'skii et~al.}(2008)\citenamefont{Yampol'skii,
  Savel'ev, and Nori}}]{Yampolskii_2008}
\bibinfo{author}{\bibfnamefont{V.~A.} \bibnamefont{Yampol'skii}},
  \bibinfo{author}{\bibfnamefont{S.}~\bibnamefont{Savel'ev}}, \bibnamefont{and}
  \bibinfo{author}{\bibfnamefont{F.}~\bibnamefont{Nori}}, \bibinfo{journal}{New
  Journal of Physics} \textbf{\bibinfo{volume}{10}}, \bibinfo{pages}{053024}
  (\bibinfo{year}{2008}).

\bibitem[{\citenamefont{Bai and Zhang}(2007)}]{Bai_2007}
\bibinfo{author}{\bibfnamefont{C.}~\bibnamefont{Bai}} \bibnamefont{and}
  \bibinfo{author}{\bibfnamefont{X.}~\bibnamefont{Zhang}},
  \bibinfo{journal}{Phys.\ Rev.\ B} \textbf{\bibinfo{volume}{76}},
  \bibinfo{pages}{075430} (\bibinfo{year}{2007}).

\bibitem[{\citenamefont{Ziegler}(2007)}]{Ziegler_2007}
\bibinfo{author}{\bibfnamefont{K.}~\bibnamefont{Ziegler}},
  \bibinfo{journal}{Phys. Rev. B} \textbf{\bibinfo{volume}{75}},
  \bibinfo{pages}{233407} (\bibinfo{year}{2007}).

\bibitem[{\citenamefont{Ryu et~al.}(2007)\citenamefont{Ryu, Mudry, Furusaki,
  and {A.W.W. Ludwig}}}]{Ryu_2007}
\bibinfo{author}{\bibfnamefont{S.}~\bibnamefont{Ryu}},
  \bibinfo{author}{\bibfnamefont{C.}~\bibnamefont{Mudry}},
  \bibinfo{author}{\bibfnamefont{A.}~\bibnamefont{Furusaki}}, \bibnamefont{and}
  \bibinfo{author}{\bibnamefont{{A.W.W. Ludwig}}}, \bibinfo{journal}{Phys. Rev.
  B} \textbf{\bibinfo{volume}{75}}, \bibinfo{pages}{205344}
  (\bibinfo{year}{2007}).

\bibitem[{\citenamefont{Ludwig et~al.}(1994)\citenamefont{Ludwig, Fisher,
  Shankar, and Grinstein}}]{Ludwig_1994}
\bibinfo{author}{\bibfnamefont{A.~W.~W.} \bibnamefont{Ludwig}},
  \bibinfo{author}{\bibfnamefont{M.~P.~A.} \bibnamefont{Fisher}},
  \bibinfo{author}{\bibfnamefont{R.}~\bibnamefont{Shankar}}, \bibnamefont{and}
  \bibinfo{author}{\bibfnamefont{G.}~\bibnamefont{Grinstein}},
  \bibinfo{journal}{Phys. Rev. B} \textbf{\bibinfo{volume}{50}},
  \bibinfo{pages}{7526} (\bibinfo{year}{1994}).

\bibitem[{\citenamefont{Tworzydlo et~al.}(2006)\citenamefont{Tworzydlo,
  Trauzettel, Titov, Rycerz, and Beenakker}}]{Tworzydlo_2006}
\bibinfo{author}{\bibfnamefont{J.}~\bibnamefont{Tworzydlo}},
  \bibinfo{author}{\bibfnamefont{B.}~\bibnamefont{Trauzettel}},
  \bibinfo{author}{\bibfnamefont{M.}~\bibnamefont{Titov}},
  \bibinfo{author}{\bibfnamefont{A.}~\bibnamefont{Rycerz}}, \bibnamefont{and}
  \bibinfo{author}{\bibfnamefont{C. W. J.}~\bibnamefont{Beenakker}},
  \bibinfo{journal}{Phys.\ Rev.\ Lett.} \textbf{\bibinfo{volume}{96}},
  \bibinfo{pages}{246802} (\bibinfo{year}{2006}).

\bibitem[{\citenamefont{{M. I. Katsnelson}}(2006)}]{Katsnelson_2006a}
\bibinfo{author}{\bibnamefont{{M. I. Katsnelson}}}, \bibinfo{journal}{Eur.
  Phys. J. B} \textbf{\bibinfo{volume}{51}}, \bibinfo{pages}{157}
  (\bibinfo{year}{2006}).

\bibitem[{\citenamefont{{Das Sarma} et~al.}(2010)\citenamefont{{Das Sarma},
  {Adam}, {Hwang}, and {Rossi}}}]{Das_Sarma_2010}
\bibinfo{author}{\bibfnamefont{S.}~\bibnamefont{{Das Sarma}}},
  \bibinfo{author}{\bibfnamefont{S.}~\bibnamefont{{Adam}}},
  \bibinfo{author}{\bibfnamefont{E.~H.} \bibnamefont{{Hwang}}},
  \bibnamefont{and} \bibinfo{author}{\bibfnamefont{E.}~\bibnamefont{{Rossi}}},
  \eprint{arXiv:1003.4731}.

\bibitem[{\citenamefont{Brey and Fertig}(2006)}]{Brey_2006b}
\bibinfo{author}{\bibfnamefont{L.}~\bibnamefont{Brey}} \bibnamefont{and}
  \bibinfo{author}{\bibfnamefont{H. A.}~\bibnamefont{Fertig}},
  \bibinfo{journal}{Phys.\ Rev.\ B} \textbf{\bibinfo{volume}{73}},
  \bibinfo{pages}{235411} (\bibinfo{year}{2006}).

\bibitem[{\citenamefont{Schomerus}(2007)}]{Schomerus_2007}
\bibinfo{author}{\bibfnamefont{H.}~\bibnamefont{Schomerus}},
  \bibinfo{journal}{Phys. Rev. B} \textbf{\bibinfo{volume}{76}},
  \bibinfo{pages}{045433} (\bibinfo{year}{2007}).

\bibitem[{\citenamefont{Blanter and Martin}(2007)}]{Blanter_2007}
\bibinfo{author}{\bibfnamefont{Y.~M.} \bibnamefont{Blanter}} \bibnamefont{and}
  \bibinfo{author}{\bibfnamefont{I.}~\bibnamefont{Martin}},
  \bibinfo{journal}{Phys. Rev. B} \textbf{\bibinfo{volume}{76}},
  \bibinfo{pages}{155433} (\bibinfo{year}{2007}).

\bibitem[{\citenamefont{Brey and Fertig}(2007)}]{Brey_2007c}
\bibinfo{author}{\bibfnamefont{L.}~\bibnamefont{Brey}} \bibnamefont{and}
  \bibinfo{author}{\bibfnamefont{H.~A.} \bibnamefont{Fertig}},
  \bibinfo{journal}{Phys. Rev. B} \textbf{\bibinfo{volume}{76}},
  \bibinfo{pages}{205435} (\bibinfo{year}{2007}).

\bibitem[{\citenamefont{Burset et~al.}(2008)\citenamefont{Burset, Yeyati, and
  Mart\'{\i}n-Rodero}}]{Burset_2008}
\bibinfo{author}{\bibfnamefont{P.}~\bibnamefont{Burset}},
  \bibinfo{author}{\bibfnamefont{A.} \bibnamefont{Levy Yeyati}},
  \bibnamefont{and}
  \bibinfo{author}{\bibfnamefont{A.}~\bibnamefont{Mart\'{\i}n-Rodero}},
  \bibinfo{journal}{Phys. Rev. B} \textbf{\bibinfo{volume}{77}},
  \bibinfo{pages}{205425} (\bibinfo{year}{2008}).

\bibitem[{\citenamefont{Burset et~al.}(2009)\citenamefont{Burset, Herrera, and
  Levy~Yeyati}}]{Burset_2009}
\bibinfo{author}{\bibfnamefont{P.}~\bibnamefont{Burset}},
  \bibinfo{author}{\bibfnamefont{W.}~\bibnamefont{Herrera}}, \bibnamefont{and}
  \bibinfo{author}{\bibfnamefont{A.}~\bibnamefont{Levy~Yeyati}},
  \bibinfo{journal}{Phys. Rev. B} \textbf{\bibinfo{volume}{80}},
  \bibinfo{pages}{041402} (\bibinfo{year}{2009}).

\bibitem[{Iye()}]{Iyengar_unpub}
\bibinfo{note}{A.P. Iyengar, Jianmin Sun, H.A. Fertig, and L. Brey,
  (unpublished).}

\bibitem[{\citenamefont{Wang et~al.}(2010)\citenamefont{Wang, Fertig, and
  Murthy}}]{Jianhui_2010}
\bibinfo{author}{\bibfnamefont{J.}~\bibnamefont{Wang}},
  \bibinfo{author}{\bibfnamefont{H.~A.} \bibnamefont{Fertig}},
  \bibnamefont{and} \bibinfo{author}{\bibfnamefont{G.}~\bibnamefont{Murthy}},
  \bibinfo{journal}{Phys. Rev. Lett.} \textbf{\bibinfo{volume}{104}},
  \bibinfo{pages}{186401} (\bibinfo{year}{2010}).

\bibitem[{\citenamefont{Wang et~al.}(2011)\citenamefont{Wang, Fertig, Murthy,
  and Brey}}]{Jianhui_2011}
\bibinfo{author}{\bibfnamefont{J.}~\bibnamefont{Wang}},
  \bibinfo{author}{\bibfnamefont{H.~A.} \bibnamefont{Fertig}},
  \bibinfo{author}{\bibfnamefont{G.}~\bibnamefont{Murthy}}, \bibnamefont{and}
  \bibinfo{author}{\bibfnamefont{L.}~\bibnamefont{Brey}},
  \bibinfo{journal}{Phys. Rev. B} \textbf{\bibinfo{volume}{83}},
  \bibinfo{pages}{035404} (\bibinfo{year}{2011}).

\end{thebibliography}
\end{document}